\documentclass[a4paper,11pt]{article}
\usepackage{pos}
\usepackage{slashed}

\usepackage{graphicx}  
\usepackage{dcolumn}   
\usepackage{bm}        
\usepackage{physics}
\usepackage{tikz}
\usepackage{mathtools}
\usepackage{amsmath} 

\newcommand{\FTEE}{FTE$^2$ }
\newcommand{\FTEEformat}{FTE$^2$}

\newcommand{\QQbar}{{Q\bar Q}}

\newcommand{\lp}{\left}
\newcommand{\rp}{\right}

\FullConference{The 42nd International Symposium on Lattice Field Theory (LATTICE2025)\\
2-8 November 2025\\
Tata Institute of Fundamental Research, Mumbai, India\\}

\title{
Topological structure of the entanglement radius of Yang-Mills flux tubes
}

\author*[a]{Rocco Amorosso}
\author[a]{Sergey Syritsyn}
\author[b,c,d]{Raju Venugopalan}

\affiliation[a]{Department of Physics and Astronomy, Stony Brook University, 
  Stony Brook, New York 11794, USA}
\affiliation[b]{Physics Department, Brookhaven National Laboratory, Upton, NY 11973, USA}
\affiliation[c]{CFNS, Department of Physics and Astronomy, Stony Brook University, Stony Brook, NY 11794, USA}
\affiliation[d]{Higgs Center for Theoretical Physics, The University of Edinburgh, Edinburgh, EH9 3FD, Scotland, UK}
\date{\today}%

\emailAdd{rocco.amorosso@stonybrook.edu}
\emailAdd{sergey.syritsyn@stonybrook.edu}
\emailAdd{raju.venugopalan@stonybrook.edu}

\abstract{
We expand on recent work~\cite{Amorosso:2026mdo}
demonstrating the existence of a novel 
entanglement radius $\xi_0$ characterizing flux tube entanglement entropy (\FTEEformat) in (2+1)D Yang-Mills theory. This physical scale 
corresponds to the intrinsic thickness of the flux tube that must be fully severed by an entangling region for color degrees of freedom in the flux tube to contribute non-zero \FTEEformat. We consider here geometries of the entanglement region $V$ on the lattice where the length of the region cross-cutting the flux tube is of the same magnitude as $\xi_0$. Our results further the conclusions of \cite{Amorosso:2026mdo} by adding detailed new information on the topological structure of the entanglement radius of color flux tubes.  

}

\FullConference{The 42nd International Symposium on Lattice Field Theory (LATTICE2025)\\
2-8 November 2025\\
Tata Institute of Fundamental Research, Mumbai, India\\}

\begin{document}
\maketitle

\section{Introduction}
Entanglement entropy is a powerful tool for understanding quantum gauge theories, revealing how information is distributed across their constituent parts.
A standard measure of entanglement is the von Neumann entanglement entropy, with 
\begin{equation}
   S^\text{EE}=-\Tr \lp(\hat{\rho}_V \log \hat{\rho}_V\rp),
\end{equation}
where $\hat{\rho}_V$ is the reduced density matrix, defined as
\begin{equation}
\label{eqn:Rho_reduced}
\hat{\rho}_V=\Tr_{\bar{V}}\hat{\rho}\,.
\end{equation}
Calculating the von Neumann entanglement entropy on the lattice is not feasible. One instead typically takes advantage of the replica trick, calculating the R\'{e}nyi entropy of order $q$,
\begin{equation}
S^{(q)}=\frac{1}{1-q}\log\lp(\Tr\hat{\rho}_V^q\rp) ,
\end{equation}
which can be calculated using a $q$-replica lattice, with $q$ an integer greater than or equal to two.
The R\'{e}nyi entropy has been used to calculate the entanglement entropy on the lattice in many recent studies \cite{Buividovich:2008kq,Buividovich:2008gq,Itou:2015cyu,Rabenstein:2018bri,Bulgarelli:2023ofi,Bulgarelli:2024onj,Bulgarelli:2024yrz,Velytsky:2008sv,Jokela:2023rba}.

However, in order to calculate the entanglement entropy in gauge theories, there are multiple well-known problems that must be addressed. Firstly, the Hilbert space of gauge invariant states cannot be factorized into gauge invariant subspaces.
As the definition of the R\'{e}nyi entanglement entropy relies on a partial trace, 
one needs a more rigorous definition of the reduced density matrix that addresses the issue of gauge invariance.
This problem has been addressed extensively in the literature, with algebraic and extended lattice constructions providing gauge-invariant definitions of $\hat{\rho}_V$.
These have been implemented in lattice calculations of the entanglement entropy using the replica method~\cite{Casini:2013rba,Ghosh:2015iwa,Buividovich:2008gq,Donnelly:2011hn,Aoki:2015bsa,Soni:2015yga,Lin:2018bud}.

Another significant issue is that the entanglement entropy in gauge field theories is typically divergent in the UV-limit, scaling with the area for $D>2$ \cite{Bombelli:1986rw,Srednicki:1993im}.
As a result, lattice studies have mainly focused on associated quantities with a finite continuum limit, an example being 
the entropic C-function~\cite{Casini:2006es,Buividovich:2008gq,Buividovich:2008kq}, which is proportional to the derivative of the entanglement entropy with respect to an infrared regulator.

In previous work, we defined a novel flux tube entanglement entropy (\FTEEformat) \cite{Amorosso:2023fzt,Amorosso:2024glf,Amorosso:2024leg,Amorosso:2025tgg,Amorosso:2026mdo}, defined as the excess entanglement entropy of a region of gluon fields that can be attributed to the presence of a quark-antiquark pair and associated flux tube in pure gauge Yang-Mills theory.
In \cite{Amorosso:2024leg,Amorosso:2024glf}, we showed that \FTEE is finite and gauge invariant.
In \cite{Amorosso:2025tgg,Amorosso:2026mdo}, we extended these studies, demonstrating that \FTEE is a powerful probe of the internal structure of the flux tube. In particular, we uncovered a novel physical scale of the theory, the entanglement radius $\xi_0$, an effective intrinsic thickness of the vibrating color flux tube.
In this work, we further explore \FTEEformat, studying the \FTEE of regions with linear size $L_x\sim\xi_0$, investigating the entangling dynamics of the color flux tube when the entangling region has length on the scale of the flux tube's internal structure.

\section{Flux Tube Entanglement Entropy (\FTEEformat) review}
In this section, we give an overview of past results regarding flux tube entanglement entropy, beginning with how it is calculated on the lattice and ending with a brief discussion of its relation to the internal structure of the color flux tube.
For a more detailed review of \FTEE in (2+1)D and (1+1)D, we refer the readers to Refs.~\cite{Amorosso:2024leg,Amorosso:2024glf}.
For detailed expositions of \FTEEformat's relevance to the internal structure of the color flux tube in (2+1)D, we refer the readers to Refs.~\cite{Amorosso:2025tgg,Amorosso:2026mdo}.

In quantum field theory, the entanglement entropy $S$ generally takes form \cite{Buividovich:2008kq,Ryu:2006bv,Ryu:2006ef,Nishioka:2006gr,Klebanov:2007ws}
\begin{equation}
\label{eq:entropyUVfinite}
S=S_{UV}+S_f\,,
\end{equation}
where $S_{UV}$ and $S_f$ are the UV-divergent and finite components of the entanglement entropy, respectively.
As noted, to extract information about $S_f$, one can study the entropic C-function.
In the presence of a static quark pair, \FTEE represents an alternative method of isolating $S_f$.
\FTEE is defined as \cite{Amorosso:2024leg}
\begin{equation}
\label{eqn:R\'{e}nyiDiff}
\tilde{S}^{(q)}_{\vert Q \bar{Q}} \equiv S^{(q)}_{\vert Q \bar{Q}}-S^{(q)} ,
\end{equation}
with $S_{\vert Q \bar{Q}}$ representing the entanglement entropy in the presence of static quarks and $S$ representing the entanglement entropy of the vacuum.
The superscript $(q)$ denotes the R\'{e}nyi entropy of order $q$, typically set to $q=2$.
\FTEE isolates the finite portion of the entanglement entropy by subtracting the UV-divergent vacuum entanglement, leaving a finite component that can be attributed to the 
flux tube.
This subtraction was demonstrated to give a finite continuum limit in (1+1)-dimensions~\cite{Amorosso:2024glf} for all gauge groups and for (2+1)-dimensions~\cite{Amorosso:2024leg,Amorosso:2025tgg} for gauge groups SU(2) and SU(3). 

\subsection{\FTEE on the lattice}
\begin{figure}[ht!]
  \centering
  \includegraphics[width=.4\textwidth]{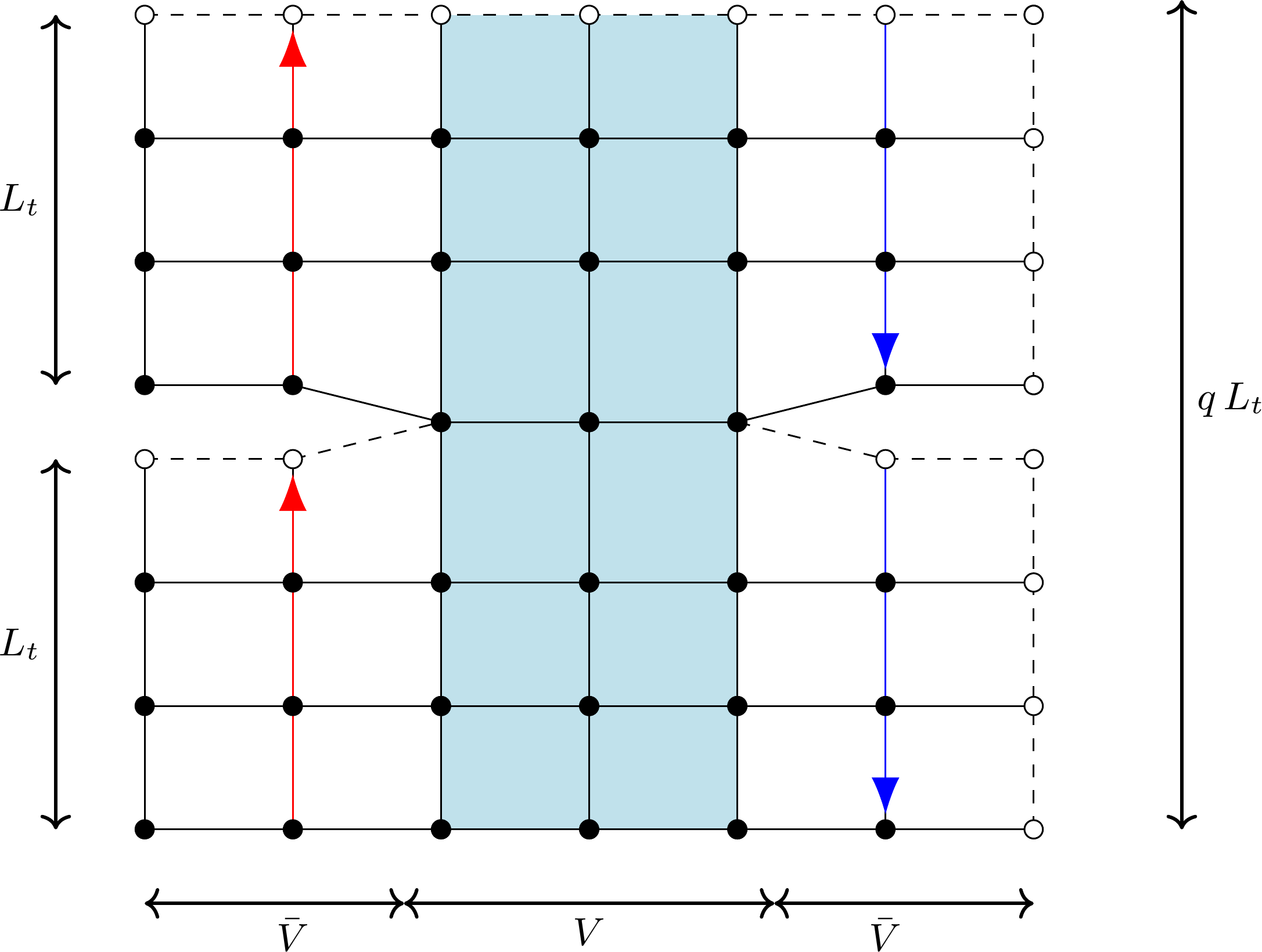}
  \caption{\label{fig:pants}
Correlator of Polyakov loops representing quark (red) and antiquark (dark blue) sources in a $q=2$ replica $L_s\times L_\tau=6\times3$ lattice with region $V$ (shaded blue) having length $L_x=2a$. 
The lattice has periodicity $L_s$ in the spatial direction. 
The lattice is $L_\tau$-periodic in region $\bar{V}$ and $2L_\tau$-periodic in region $V$. 
Dashed gauge links and open vertices are images, respectively, of solid gauge links
and filled vertices, determined by the respective temporal or spatial periodic boundary conditions.
}
\end{figure}

To calculate \FTEE on the lattice, one utilizes Polyakov loops and the replica trick.
For a detailed explanation of the lattice structure involved, we refer the reader to Ref.~\cite{Amorosso:2024leg}. 
We follow the procedure of
Refs.~\cite{Itou:2015cyu,Rabenstein:2018bri,Calabrese:2004eu} to generate $\hat{\rho}_V^q$, stacking $q$ independent replicas, each of size $L_\tau=\beta=T^{-1}$ in the temporal direction.
The resulting lattice has two regions: region $V$ and its complement $\bar{V}$, each with a different temporal periodicity.
In region $V$, gauge links from adjacent replicas, representing in- and out-states of $\hat{\rho}_V$, are identified, resulting in a long periodic time extent $qL_\tau$.
Explicitly, for gauge links $U$ oriented in spatial direction $\hat{\mu}$, we identify $U_{\hat{\mu}}(\tau=\beta)^{(r)}=U_{\hat{\mu}}(\tau=0)^{(r+1)}$ for $r<q$ and   $U_{\hat{\mu}}(\tau=\beta)^{(q)}=U_{\hat{\mu}}(\tau=0)^{(1)}$, where superscript $(r)$ denotes the replica number.
In region $\bar{V}$, gauge links within the same replica at $\tau=0$ and $\tau=\beta$ are identified, representing the in- and out-states of $\hat{\rho}$ in $\bar{V}$, which must be traced over to give the reduced density matrix of region $V$.
Explicitly, $U_{\hat{\mu}}(\tau=0)^{(r)}=U_{\hat{\mu}}(\tau=\beta)^{(r)}$ for each replica $r$, resulting in normal periodic time extent $L_\tau$.
This geometry is illustrated in (1+1)-dimensions in Fig.~\ref{fig:pants}.

Quark and antiquark sources, represented by temporal Polyakov loops on the lattice, are placed in $\bar{V}$,
resulting in $2q$ Polyakov loops of length $L_\tau$.
\FTEE can be calculated from the correlation of these $2q$ Polyakov loops~\cite{Amorosso:2024leg,Amorosso:2023fzt},
\begin{equation}
    \tilde{S}^{(q)}_{\vert\QQbar}\equiv S^{(q)}_{\vert\QQbar} - S^{(q)}
=-\frac{1}{q-1}\left(\ln \frac{Z^{(q)}_{\vert\QQbar}}{Z^{q}_{\vert\QQbar}}-\ln \frac{Z^{(q)}}{Z^q}\right)
=-\frac{1}{q-1}\ln\left( \frac{\langle\prod\limits_{r=1}^q\text{Tr}P_{\vec{x}}^{(r)}\text{Tr}P^{(r)\dagger}_{\vec{y}}\rangle}{(\langle\text{Tr}P_{\vec{x}}\text{Tr}P^\dagger_{\vec{y}}\rangle)^q}\right)\,.
\end{equation}
where $\Tr P_{\vec{x}}^{(r)}$ is the trace of the Polyakov loop at location $\vec{x}$ in replica $r$,
while $\Tr P_{\vec{x}}$ is the trace of the Polyakov loop in the standard one-replica lattice.

\subsection{Previous results
  \label{sec:models}}
\FTEE has been studied in detail in both (1+1) and (2+1) dimensions.
The (2+1)D studies of \FTEE have primarily utilized the ``half-slab'' geometry depicted in Fig.~\ref{fig:refinedsmallslab}(left), where both the quark and the antiquark were placed in region $\bar{V}$ and separated by a distance $L$.
Through studying the half-slab geometry, \FTEE was found to split into two components
\begin{equation}
\label{eq:vib+int}
\tilde{S}^{(q)}_{\vert Q \bar{Q}} = S_\text{internal} + S_\text{vibrational}\,,
\end{equation}
where $S_\text{internal}$ represents the entanglement entropy due to color degrees of freedom within the flux tube, while $S_{\text{vibrational}}$ represents the entanglement due to mechanical vibrations of the flux tube.
In this work, we focus primarily on the internal entropy, which dominates $\tilde{S}^{(q)}_{\vert Q \bar{Q}}$ at the quark separations studied here ($0.67<L\sqrt{\sigma_0}<0.89$), and is known to make up $\sim95\%$ of \FTEE~\cite{Amorosso:2024leg}.\footnote{$S_{\text{vibrational}}$ scales $\propto\frac{1}{4}\log(L)$ at $q=2$. Therefore, for sufficiently long quark separations, the vibrational entanglement entropy will be the dominant contribution.}

Motivated by the results of \FTEE in (1+1)D, $S_{\rm internal}$ was conjectured to take form \cite{Amorosso:2024leg,Amorosso:2024glf}
\begin{equation}
\label{eq:intersectionProbability}
S_\text{internal}=\langle F\rangle \cdot \log(N_c)\,,
\end{equation}
where $\langle F\rangle$ is the average number of times the flux tube crosses the $V/\bar V$ spatial boundary.
We obtain results for (2+1)D SU(2) Yang-Mills
consistent with this expectation~\cite{Amorosso:2024leg}.
When region $V$ is placed between the quark pair, forcing the flux tube between them to cross region $V$, then $\tilde{S}_{\QQbar}\sim2\ln(2)$ due to  crossing the $V/\bar{V}$ boundary twice.
When the entangling region was moved far from the quark pair, $\tilde{S}_{\QQbar}\sim0$, as it is energetically unfavorable for the flux tube to deflect long distances and cross into region $V$.
Moving the slab $V$ such that it only partially separates quark and antiquark produced an interpolation between these two limits consistent with Gaussian deflection of the effective string.

This study was extended to different values of $N_c$ and regions $V$ composed of multiple disconnected subregions (allowing $F>2$) in Ref.~\cite{Amorosso:2025tgg}, 
finding agreement with the $\langle F\rangle\ln N_c$ with small but significant deviations.
By studying multi-slab geometries, these deviations are understood to be caused by a small finite thickness of the vibrating flux tube.
In \cite{Amorosso:2025tgg,Amorosso:2026mdo}, the \FTEE of a region $V$ consisting of two disconnected subregions was studied as a function of the distance between the two subregions.
\FTEE  varies smoothly as a function of the separation before jumping sharply as the two subregions are brought into contact.
This result is incompatible with a zero-width  effective string. It instead is compatible with a thick vibrating effective string that must be \emph{fully severed} in order to contribute $F\ln N_c$ to \FTEE.
This novel model of a vibrating effective string was studied using multiple geometries of region $V$ in \cite{Amorosso:2026mdo}, giving a consistent value of the effective thickness of the flux tube. This physical quantity, which we call ``entanglement radius'', has the value $\xi_0=0.185(6)\sigma_0^{-1/2}$, where $\sigma_0$ is the string tension at zero temperature.

This entanglement radius is related to another scale of the flux tube's internal structure, the intrinsic width \cite{Caselle:2012rp,Caselle:2026coc}.
Taking the derivative of \FTEE with respect to the displacement $x$ of the entangling region from the quark-antiquark pair line, $\partial_x\tilde{S}_{\vert\QQbar}$ was found to have a Gaussian peak centered at $x=-\xi_0$ and exponential decay $\propto e^{-x/\lambda}$ in its tails -- this decay length $\lambda\sqrt{\sigma_0}=0.223(15)$ is consistent with both the inverse glueball mass~\cite{Teper:1998te} 
and the intrinsic width of the flux tube~\cite{Caselle:2026coc}.
This deviation from Gaussianity due to the intrinsic width suggests that the entanglement radius is not fixed at one value, but instead follows a distribution.
In \cite{Amorosso:2026mdo}, it was proposed that the entanglement radius arises from a distribution $P(\xi)$ with mean $\xi_0$ and support over a broad range of values $\xi$.
In this work, we explicitly test this proposal, studying regions $V$ with length $L_x\sim\xi_0$.

\section{Results for small-slab geometry}
\label{sec:results}
\begin{figure}[t!]
\centering
\includegraphics[width=.35\textwidth]{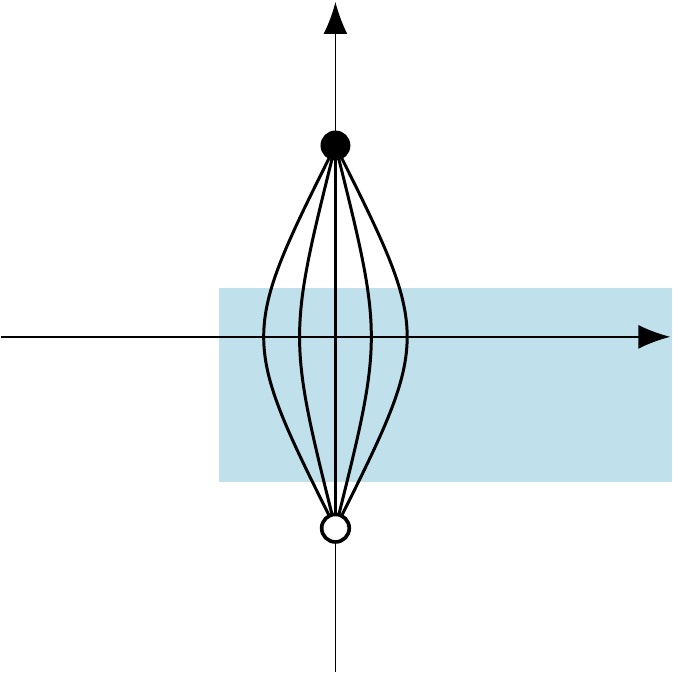}
\hspace{2cm}
\includegraphics[width=.35\textwidth]{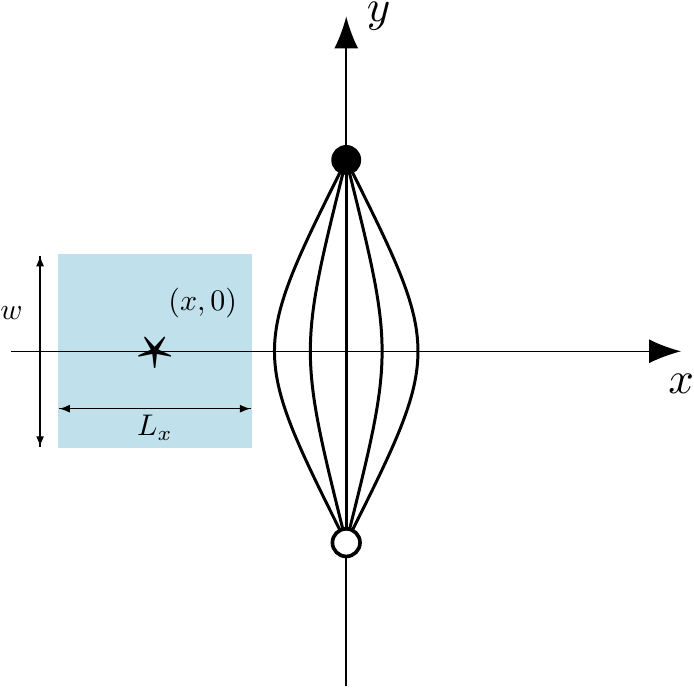}
\hfill
\\
\caption{\label{fig:refinedsmallslab}
(Left) A depiction of the ``half-slab'' geometry of Ref.~\cite{Amorosso:2024leg}, with region $V$ (shaded blue) having length $L_x$ equal to half of the $x$-extent of the lattice.
(Right) A depiction of the ``small-slab'' geometry, with region $V$ (shaded blue) of variable length $L_x$ in the $x$ direction.
The quark and antiquark sources are separated by distance $L$ and located at $(0,\pm L/2)$.
}
\end{figure}

\begin{figure}[ht!]
\centering
\includegraphics[height=.2\textwidth]{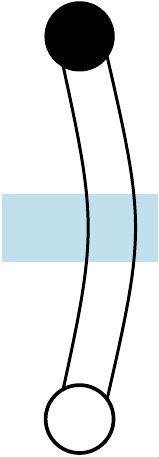}
\hspace{3cm}
\includegraphics[height=.2\textwidth]{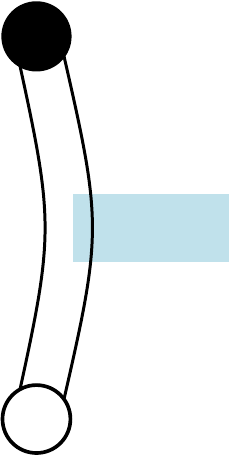}
\hspace{3cm}
\includegraphics[height=.2\textwidth]{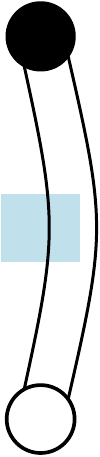}
\caption{\label{fig:schemCenter} 
Examples of different overlaps of the flux tube with the entangling region $V$ (shaded blue). (Left) The flux tube exhibits a full boundary crossing and contributes $2\ln 2$ to \FTEE. (Center) The center of the slab $V$ is too far to the right ($x$ too large) to fully sever the flux tube; the flux tube exhibits a partial boundary crossing and does not contribute to \FTEE. (Right) The length ($L_x$) of the slab $V$ is too small relative to $\xi_0$; the flux tube exhibits a partial boundary crossing and does not contribute to \FTEE.
}
\end{figure}

We present here 
calculations further exploring the structure of \FTEE for $SU(2)$ Yang-Mills gauge theory on a (2+1)D lattice.
We utilize the standard Wilson plaquette Yang-Mills action, performing Monte Carlo sweeps with details identical to those of \cite{Amorosso:2024leg} on a $128\times256\times64$ lattice.
Simulations are performed at $T_c/4$ and lattice coupling $\beta=24.744$, consistent with lattice spacing $a\sqrt{\sigma_0}=0.0557(11)$.
\FTEE values 
were obtained for the ``small-slab'' geometry configuration shown in Fig.~\ref{fig:refinedsmallslab}(right), with slab length $L_x\sim\xi_0$ and width $w=4a$.
The interpretation of these results depends strongly on the observation in  \cite{Amorosso:2025tgg,Amorosso:2026mdo} that \FTEE behaves topologically, such that the flux tube has to be \emph{completely severed by entangling region $V$}, crossing the boundary with its full width, in order to contribute to \FTEEformat.
Fig.~\ref{fig:schemCenter} illustrates flux tube configurations that overlap with region $V$ to create both full and partial boundary crossings.

\begin{figure}[t!]
\centering
\includegraphics[width=.45\textwidth]{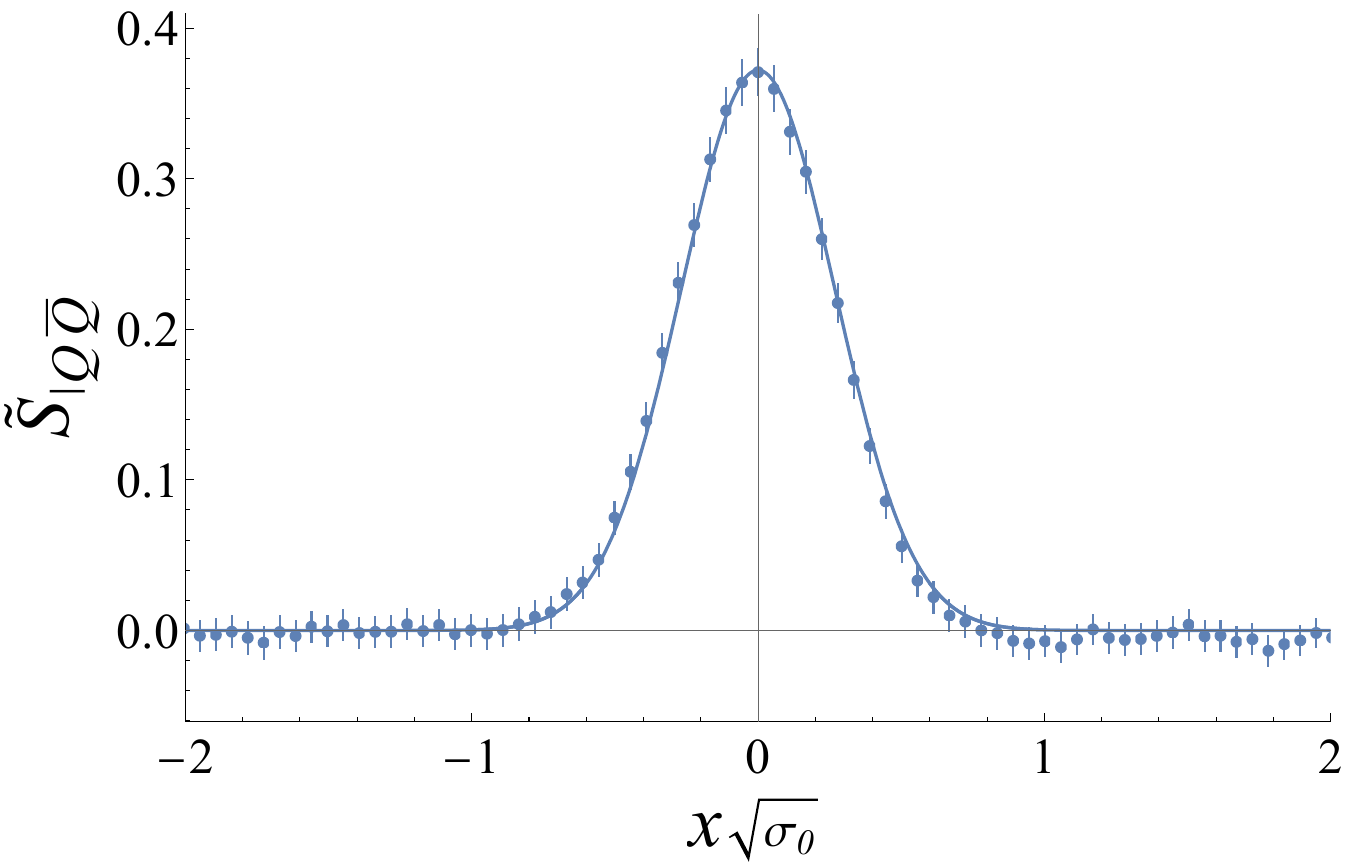}
\hfill
\includegraphics[width=.45\textwidth]{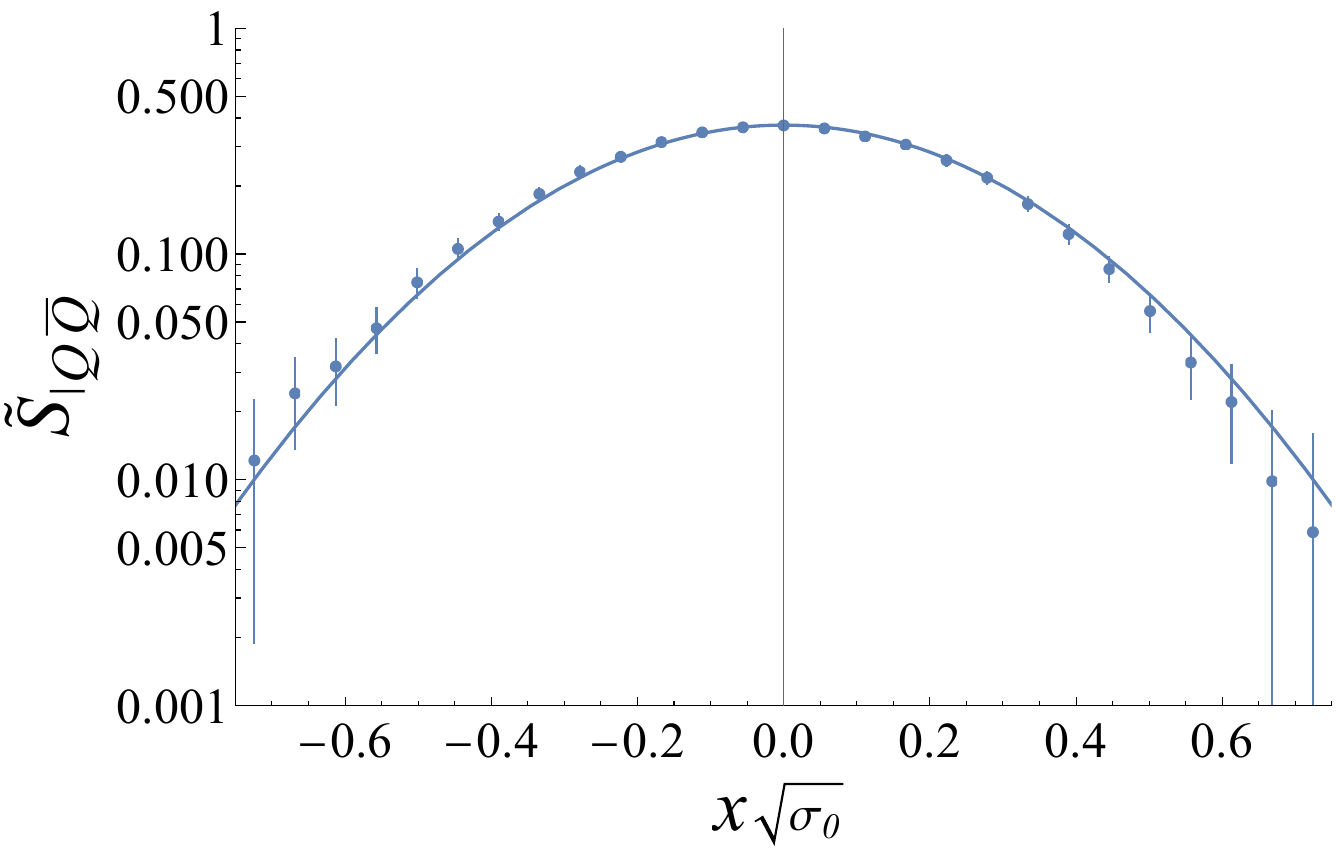}
\\
\caption{\label{fig:GaussSmall} 
Dependence of \FTEE $\tilde{S}_{\vert\QQbar}$ on the transverse position of the slab $x$ computed with small-slab geometry (Fig.~\ref{fig:refinedsmallslab}), in linear (left) and logarithmic (right) scale. 
Calculations are done in (2+1)D $SU(2)$ Yang-Mills theory with $\QQbar$ distance fixed to $L\sqrt{\sigma_0}\approx0.67$.
The solid line shows the fit to a Gaussian shape.
}
\end{figure}

\begin{figure}[t!]
\centering
\includegraphics[width=.45\textwidth]{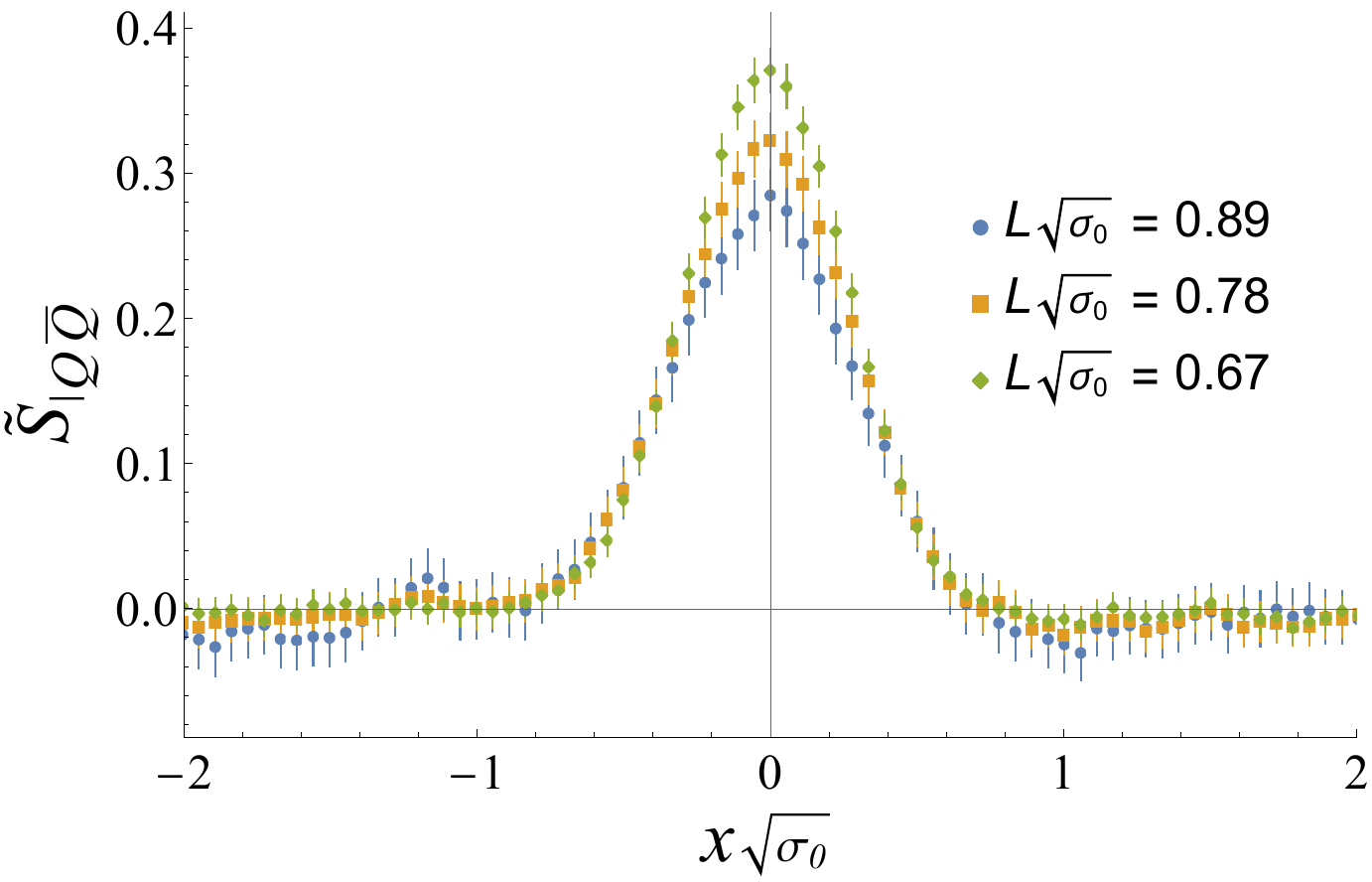}
\hfill
\includegraphics[width=.45\textwidth]{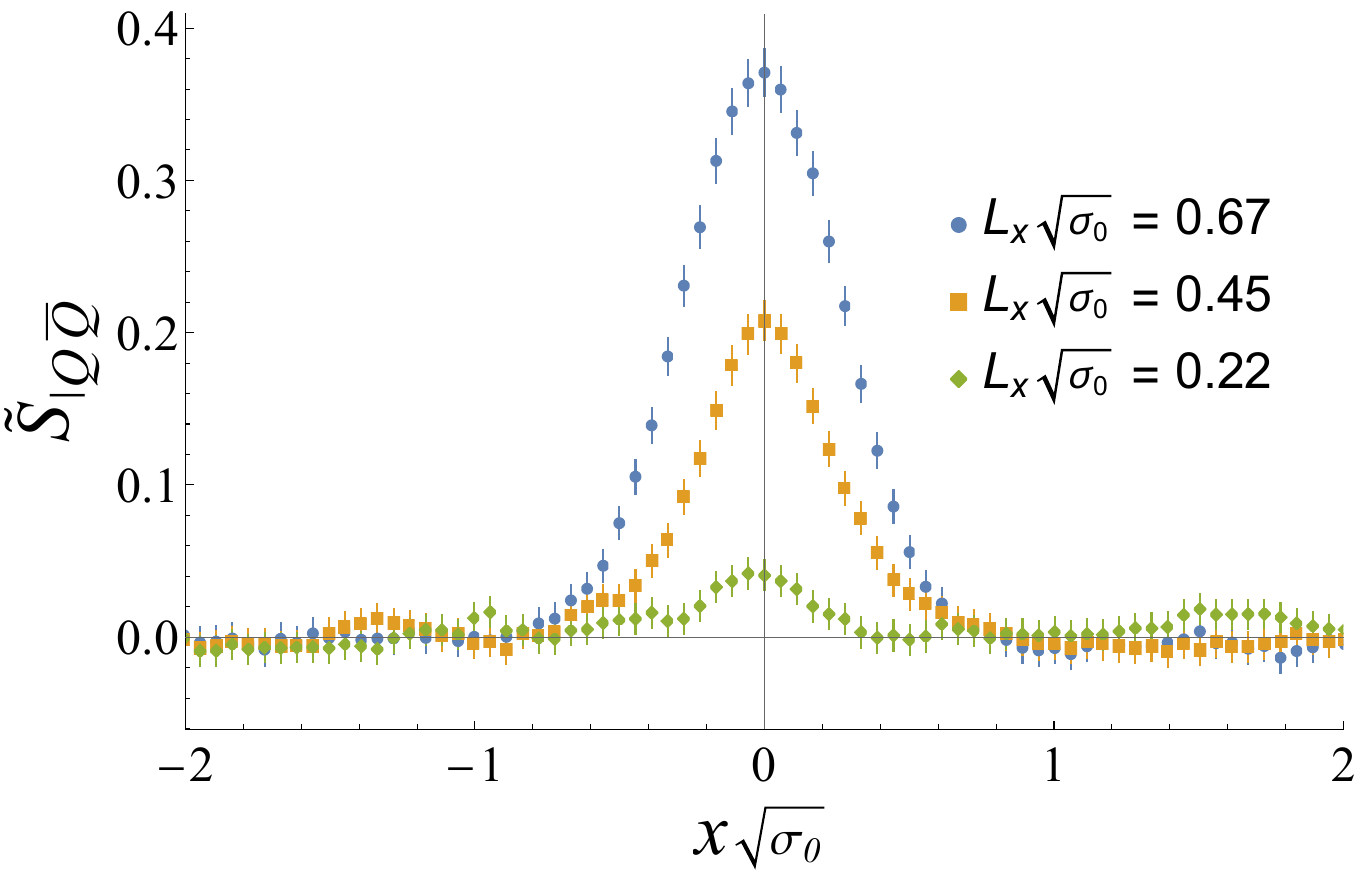}\\
\caption{\label{fig:LdepSmall}
(Left)  Plot of \FTEE as a function of $x$ with varying values of the quark separation $L$. 
Note that the \FTEE value of the peak of each profile decreases as $L$ increases.
(Right)
\FTEE as a function of $x$ with varying values of the slab length $L_x$. 
The \FTEE value of the peak of each profile increases significantly as $L_x$ increases.
At the other extreme $L_x\sqrt{\sigma_0}=0.22$, the \FTEE contribution at $x=0$ is still non-zero despite $L_x < 2\xi_0\sim0.37\sigma_0^{-1/2}$.
}
\end{figure}

We begin by studying the $x$-dependence of \FTEEformat.
The $x$-dependence of \FTEE with a quark separation of $L\sqrt{\sigma_0}=0.67$ is shown in Fig.~\ref{fig:GaussSmall}. 
\FTEE has a roughly Gaussian form as a function of $x$, in line with expectations of a thick effective string exhibiting Gaussian deflection with $L_x\sim\xi_0$.
We expect the mean square transverse deflection of the flux tube to grow logarithmically~\cite{Luscher:1980iy} as the quark-antiquark separation is increased, resulting in a wider \FTEE profile with a less-pronounced peak.
The dependence of \FTEE on the quark-pair separation $L$ is shown in Fig.~\ref{fig:LdepSmall}(left).
As expected, the Gaussian peak becomes significantly smaller and wider as the quark separation increases.
While we expect larger quark separations to produce a profile with higher tails, we lack statistical precision to resolve such features in sufficient detail.

We expect \FTEE to grow as a function of $L_x$.
Comparing Fig.~\ref{fig:schemCenter} left and right, qualitatively, the shorter slab is less likely to exhibit a full boundary crossing than a longer slab.
To quantify these remarks, consider a region $V$ centered at $x$ with length $L_x$.
Also consider an effective string representing the flux tube with entanglement radius $\xi$.
The center of the effective string deflects to point $(\chi,0)$, with its leftmost and rightmost walls being at $\chi_L=\chi-\xi$ and $\chi_R=\chi+\xi$ respectively.
The flux tube, and region $V$, exhibit a full boundary crossing when the entirety of the flux tube is in $V$.
This is equivalent to satisfying both conditions $(x-L_x/2) < \chi_L$, $\chi_R<(x+L_x/2)$.
Increasing $L_x$ relaxes these bounds, allowing flux tubes with a wider range of deflections $\chi$ to contribute \FTEE as well.
We do in fact observe this expected behavior, with \FTEE strongly depending on $L_x$.
The results of this study can be seen in Fig.~\ref{fig:LdepSmall}(right).

The $L_x$-dependence of \FTEE allows us to probe the entanglement radius further.
In Ref.~\cite{Amorosso:2026mdo}, the average entanglement radius was found to be $\xi_0=0.185(6)\sigma_0^{-1/2}$.
There it was also proposed that the entanglement radius of the flux tube follows some distribution $P(\xi)$.
Studying the $L_x$-dependence of \FTEE gives us explicit confirmation that $P(\xi)$ has support over values $\xi<\xi_0$.
Indeed, we observe that \FTEE is nonzero even for $L_x<2\xi_0$. Specifically, at both $L_x\sqrt{\sigma_0}=0.22$ and $L_x\sqrt{\sigma_0}=0.33$ we observe \FTEE values significantly above zero despite their slab length being less than $2\xi_0\sqrt{\sigma_0}=0.370(12)$.

\begin{figure}[t!]
\centering
\includegraphics[width=.42\textwidth]{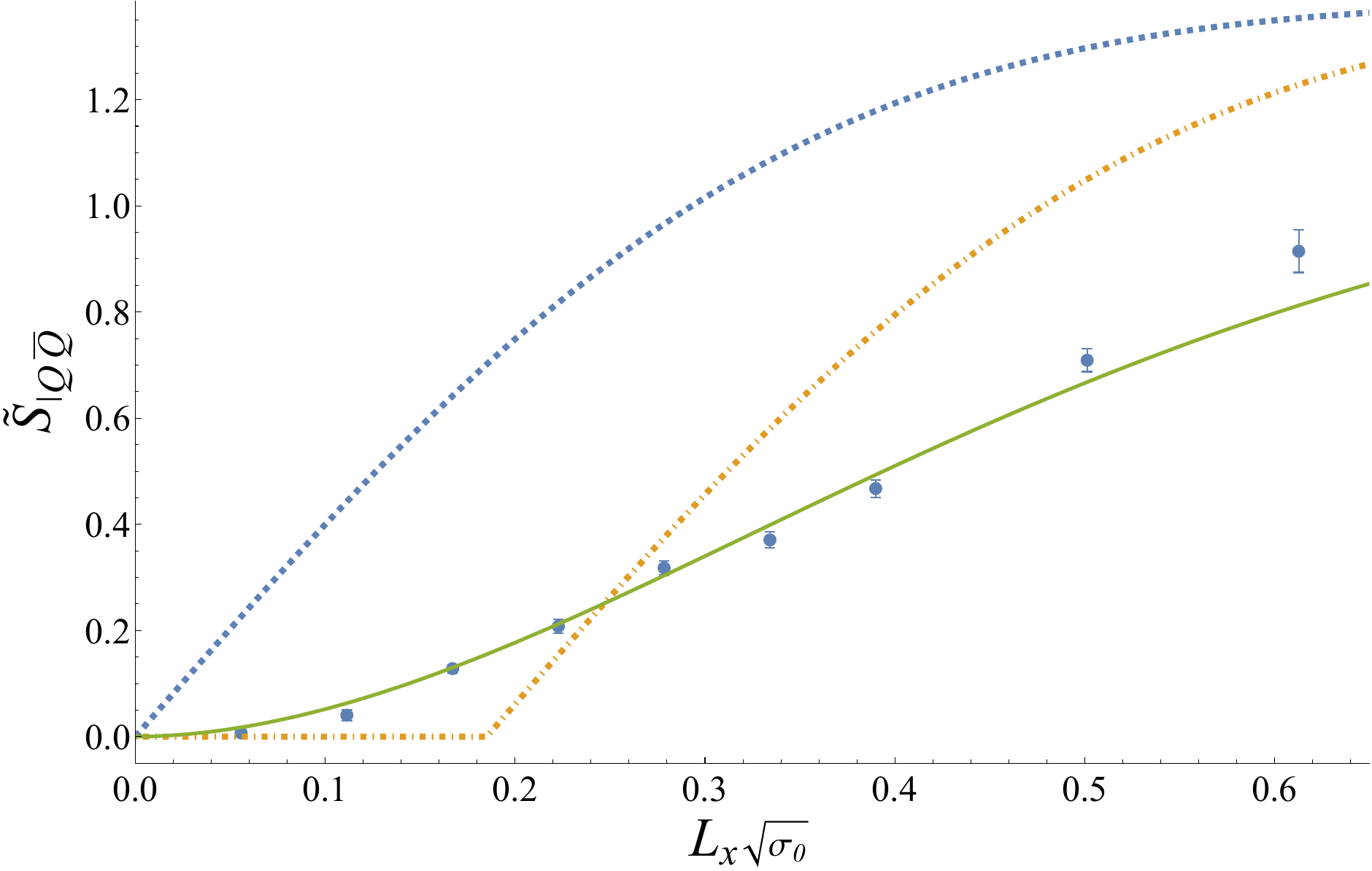}\\
\caption{\label{fig:modelsSmall}
\FTEE at $x=0$ as a function of the slab length $L_x$ with 
fixed $\QQbar$ pair separation $L\sqrt{\sigma_0}=0.67$.
The data is compared to three models of an effective string with zero entanglement radius (dashed, blue), with a fixed entanglement radius $\xi=\xi_0$ (dashed-dotted, orange), and with varying exponentially-distributed entanglement radius $\sim P(\xi)$ (solid, green).
All models assume Gaussian probability for the transverse deflection of the flux tube with standard deviation $s\sqrt{\sigma_0}=0.27$.
}
\end{figure}

Comparing the \FTEE values at $x=0$ for varying slab lengths, one can examine different models of flux tube entanglement.
In Fig.~\ref{fig:modelsSmall}, we show \FTEE as a function of $L_x$ alongside the expectations of three models: an effective string with zero entanglement radius, a fixed entanglement radius $\xi=\xi_0$, and an exponentially decaying distribution $P(\xi)$.
One can see that the fixed entanglement radius model does not describe the data well; on the contrary, the data shows stronger agreement with the model in which the entanglement radius is allowed to follow an exponential distribution.

\section{Discussion}
In this work, we continued our studies of the color flux tube's internal dynamics in SU(2) (2+1)D Yang-Mills theory using flux tube entanglement entropy.
By studying the \FTEE with spatial regions having length comparable to the entanglement radius, a novel scale parameterizing the flux tube's intrinsic thickness~\cite{Amorosso:2026mdo}, we were able to obtain preliminary results detailing the internal structure of the color flux tube.
We found that \FTEE behaves largely as expected, having a roughly Gaussian form as a function of slab transverse position, decreasing as quark and antiquark are further separated, and increasing with the length of the entangling region.
These preliminary results support the conclusions of Ref.~\cite{Amorosso:2026mdo} that the flux tube is well-modeled by an effective string with Gaussian deflection and finite non-zero entanglement radius.
Our results also strongly suggest that the entanglement radius follows a distribution $P(\xi)$ with mean $\xi_0$ and support over a wide distribution of values $\xi$.
In the future, we expect to increase statistical precision for our $SU(2)$ results as well as extend our studies of the small-slab geometry to larger $N_c$.
With these additional data, we expect to make quantitative statements regarding the flux tube's internal structure and the functional form of the distribution of entanglement radii $P(\xi)$.

\section*{Acknowledgements} 
R.A. is supported by the Simons Foundation under Award number 994318 (Simons Collaboration on Confinement and QCD
Strings). S.S. and R.A. (partially) are supported by the National Science Foundation under award PHY-2412963. 
In addition, R.A. is supported in part by the Office of Science, Office of Nuclear Physics,
 U.S. Department of Energy under Contract No. DEFG88ER41450. R.V. is supported by the U.S. Department of Energy, Office of Science under contract DE-SC0012704. R.V.'s work on quantum information science is supported by the U.S. Department of Energy, Office of Science, National
 Quantum Information Science Research Centers, Co-design Center for Quantum Advantage (C$^2$QA) under contract number  DE-SC0012704. R.V. was also supported at Stony Brook by the Simons Foundation as a co-PI under Award number 994318 (Simons Collaboration on Confinement and QCD Strings). R.V. acknowledges support from the Royal Society Wolfson Foundation Visiting Fellowship and the hospitality of the Higgs Center at the University of Edinburgh. 
The authors thank Stony Brook Research Computing and Cyberinfrastructure and the Institute for Advanced Computational
Science at Stony Brook University for access to the Seawulf HPC system, which was made possible by grants from the
National Science Foundation (awards 1531492 and 2215987) and matching funds from the Empire State Development’s Division
of Science, Technology and Innovation (NYSTAR) program (contract C210148).

\bibliography{entent-lat23proc}
\bibliographystyle{JHEP}

\end{document}